\documentclass[useAMS]{mn2e}
\usepackage{color}
\def\deg{{$^0$}}
\def\sdeg{{deg$^2$}}

\def\apj{{\it Astrophys.~J.}}  %Astrophysical Journal%
\def\apjs{{\it Astophys.~J.~Suppl.}}  %Astrophysical Journal Supplements%
\def\mnras{{\it Mon. Not. R. astr. Soc.}}      %Monthly Notices of the Royal%
\def\aap{{\it Astron.~Astrophys.}}     %Astronomy & Astrophysics%
\def\aj{{\it Astron.~J.}} %Astronomical Journal

\definecolor{Blue}{rgb}{0.3,0.3,0.9}
\definecolor{Red}{rgb}{0.9,0.3,0.3}

\title[Survey at 43 GHz]{A 43-GHz VLA Survey in the ELAIS N2 Area}
\author[J. V. Wall et al.]{J. V. Wall$^{1}$\thanks{E-mail:
jvw@phas.ubc.ca}, R. Perley$^2$, R. A. Laing$^3$, S. Stotyn$^{1,4}, $Angela C. Taylor$^5$, and J. Silk$^5$\\
$^{1}$Department of Physics and Astronomy, University of British Columbia, 6224
Agricultural Road, Vancouver, V6T\,1Z1, Canada\\
$^2$ NRAO Array Operations Center, P.O. Box 0, Socorro, NM, 87801-0387, USA\\
$^3$ ESO, Karl-Schwarzschild-Strasse 2, D-85748 Garching-bei-M\"{u}nchen, Germany\\
$^4$ now at Department of Physics and Astronomy, University of Waterloo, Waterloo, ON, N2L\,3G1, Canada\\
$^5$ Astrophysics Group, Department of Physics, University of Oxford, Keble Road, Oxford, OX1\,3RH, UK}

\begin{document}

\setlength{\topmargin}{-16mm}

\date{2010 Feb 7}

%\pagerange{\pageref{firstpage}--\pageref{lastpage}} \pubyear{2009}

\maketitle

%\label{firstpage}

\begin{abstract}

We describe a survey in the ELAIS N2 region with the VLA at 43.4~GHz, carried out with 1627 independent 
snapshot observations in D-configuration and covering about 0.5 \sdeg. One certain source is detected, 
a previously-catalogued flat-spectrum QSO at z=2.2. A few ($<5$) other sources may be present at about 
the $3\sigma$ detection level, as 
determined from positions of source-like deflections coinciding with blue stellar objects, or with 
sources from lower-frequency surveys.  Independently we show how all the source-like
detections identified in the data can be used with a maximum-likelihood
technique to constrain the 43-GHz source counts at a level of 
$\sim$7~mJy. Previous estimates of the counts at 43 GHz, based on 
lower-frequency counts and spectral measurements, are consistent with these constraints, although
the present results are suggestive of somewhat higher surface densities at the 7~mJy level. 
They do not provide direct evidence of intrusion of a previously unknown source population, although the
several candidate sources need examination before such a population can be ruled out.

\end{abstract}


\section{Introduction}

We present a survey to detect extragalactic radio sources at 43 GHz with repeated independent pointings 
of the D-configuration VLA in snapshot mode.  The survey is centered on ELAIS-N2 (RA~16h~36m~48s, Dec 
+41\deg~01'~45'', J2000), and covers about 0.5 \sdeg~down to 7~mJy. The primary objective was to 
establish the 43-GHz source surface density at mJy levels, an issue of importance in assessing the 
discrete-source foreground contamination of the CMB signal.

It is only recently that we have gained insight into what populates the radio sky at survey frequencies 
above 5~GHz. By this frequency some 60 per cent of the sources at Jy levels have `flat' spectra ($\alpha > 
-0.5$, where flux density $S$ is related to frequency $\nu$ via $S \propto \nu^\alpha$) over some 
region of the radio domain. This is the signature of optically-thick synchrotron emission and compact 
structure, so-called `flat-spectrum' sources. (Surveys at frequencies above 150~GHz do find
inverted-spectrum sources presumably dominated by free-free emission, probably dust emitters related 
to or part of the sub-mm galaxy population. These do not appear at frequencies as low as 43~GHz -- 
see Viera et al. 2009). `Flat-spectrum' objects do not generally have $\alpha = 0$ over a large frequency 
range; rather their spectra show inversions, points of inflection, and sometimes single peaks (e.g. 
Gigahertz-Peaked Spectra or GPS sources). The optical counterparts of the great majority of these 
sources are stellar in appearance and are Flat-Spectrum Radio QSOs (FSRQs) or BL\,Lac objects.
 
How rapidly does this trend to such objects and away from conventional steep-spectrum -- generally 
optically-thin spectrum radio galaxies -- continue with increasing frequency above 5~GHz? When and at 
what flux-density level do we find new populations of flat or inverted-spectrum objects?

The pioneering survey of Brandie \& Bridle (1974) covered 0.8~sterad at 8~GHz and found 55 sources, 
of which ``more than 2/3'' were stellar objects, i.e. QSOs or BL\,Lac objects. Subsequent deep surveys by 
Windhorst et al. (1993) enabled 
definition of an 8.4-GHz source count down to 10~$\mu$Jy.  Brandie and Bridle found that at their 
relatively high flux densities, 70 per~cent of the sources have flat or inverted spectra. Some 
spectrally-extreme sources certainly exist -- Edge et al. (1998) found two with spectral 
peaks at $\ge$~20~GHz -- but the proportion of such objects to be detected in very high frequency surveys 
remains unknown. 

The first ventures at higher frequencies were to search for contaminants to CMB experiments, with the 
Ryle Telescope  survey at 15~GHz leading the way. Taylor et al. (2001) found 66 sources in 63 \sdeg~above 
20~mJy. The continuation of this came to be the Cambridge 9C survey (Waldram et al. 2003; Bolton et al. 
2004; Waldram et al. 2009). Some 
465 sources were catalogued above 25 mJy in 520 \sdeg. Examination of a subset of these sources 
(Bolton et al. 2004) revealed that (a) at a flux limit of 25~mJy 18 per cent show spectra that peak 
at 5 GHz or above; and (b) at 60~mJy 27 per cent peak at 5~GHz or above, and increasing the flux 
limit increases the fraction of rising-spectrum sources.  (The statistical dependence of radio 
spectrum on flux-density level has been discovered relatively regularly since 1970; see e.g. 
Kellermann \& Wall 1987.) Colours and the 
stellar nature of optical counterparts indicate that between a third and a half of the peaked-spectrum 
sources are QSOs.

The major contribution to surveys and samples at these higher frequencies has come 
from AT20G, the large-area 20-GHz survey with the Australia Telescope Compact Array (Ricci et al 2004; 
Sadler et al. 2006; Massardi et al. 2008). By 2007 the survey had 
covered $-90$\deg\ $ < \delta < -15$\deg\ cataloguing 4400 sources
down to $\sim$50~mJy; another 1500 are expected in the final survey zone $-15$\deg$ < \delta < 0$\deg. 
These catalogues will render incorrect the statements made at the start of each of these papers 
about our poor knowledge of the high-frequency radio sky. Massardi et al. studied the 320 brightest 
sources from the 2007 version of the catalogue. The major conclusions do not differ dramatically  
from those from 9C: (a) most sources do not show power-law spectra; (b) the spectral complexity 
ensures that it is impossible to select a low-frequency sample which will constitute a complete 
high-frequency sample; (c) spectral steepening is common and correlates with redshift, perhaps 
due to changing rest frame frequency; (d) 77 per cent of the brightest-source sample are QSOs, 
19 per cent are galaxies, and the remainder are blank on the SuperCosmos Sky Survey (SSS; see
{\it www-wfau.roe.ac.uk/sss/}) scanned 
versions of the UKSTU plates. The galaxies have a median $B_J$ = 17.7, the QSOs 18.6 -- these 
medians lie far above the plate limits. 

Further characterization of the source population at high frequencies has come from multi-frequency
flux-density measurements of sources detected in the WMAP (Wilkinson Microwave Anisotropy Probe) CMB
surveys. WMAP has produced complete samples of the brightest radio 
sources over the sky to a flux-density limit of about 2 Jy at each of its primary frequencies 
23, 33, 43, 61 and 94 GHz (Hinshaw et al. 2007; L\'{o}pez-Caniego et al. 2007; Wright et al. 2009). 
With measurements at 16 and 33 GHz using the VSA 
and AMI, Davies et al. (2009) and Franzen et al. (2009) found the following for a sample complete 
to 1.1~Jy at 33 GHz: 
(1) the proportion of flat-spectrum objects is very high: 93 per cent 
have spectra with $\alpha^{34}_{14~\rm{GHz}} > -0.5$; (2) 44 per cent have 
$\alpha^{34}_{14~\rm{GHz}} > 0.0$,
i.e. rising spectra; (3) the variable-source bias (a selection effect which mimics an Eddington bias; 
see Wall et al. 2005; Wall 2007) is
in clear evidence from comparing repeated WMAP and AMI 33-GHz flux densities; (3) on timescales of
$\sim$1.5yr, 20 per cent of sources varied by more than 20 per cent at 33 GHz with a high degree of
correlation between variation at 16 and 33 GHz; and (4) variable sources have a marginally flatter
spectrum ($\overline{\alpha} = -0.06\pm0.05$) than non-variable sources 
($\overline{\alpha} = -0.13\pm0.04$).

Sadler et al. (2008) attempted to characterize the populations to be found at yet higher frequencies.
With flux densities measured for 130 sources at 95 GHz and 20~GHz with the ATCA, the
resultant distribution of spectral indices together with source counts from AT20G was used to 
predict the source count at 95 GHz above a level of 100 mJy. This prediction is lower than 
that from models of luminosity functions and evolution (De Zotti et al. 2005). The
authors conclude that this is due to the fact that the majority of sources with flat or 
rising spectra over the range 5 to 20~GHz show a spectral turnover between 20 and 95~GHz. 
%This paper is of importance in the present investigation and we return to it below.

It is the purpose of this paper to describe a limited survey at mJy levels at 43.4 GHz 
with the VLA, and to use this to construct a source count at this frequency. We want 
to see if there is evidence for emergent new populations that might affect future CMB experiments 
such as Planck. In addition we wish to track the space density of the different classes of object 
which emerge as survey frequency is raised, and for this, a source count is an essential complement 
to radio and optical spectral data. Descriptions of this space density for {\em combined} 
populations of flat-spectrum objects have been obtained (e.g De Zotti et al. 2005; Wall et al. 2005; 
Ricci et al. 2006), but not for the individual (e.g. GPS) populations.

With regard to a source count at 43 GHz, the WMAP all-sky survey at 43~GHz defined the 
high-flux-density end of the count as well as it can ever be done. To extend the 43-GHz count 
to lower flux levels, Waldram et al. (2007) used their 15-GHz 9C survey plus spectral information 
to predict source counts at this frequency (as well as at several other frequencies). At a somewhat
lower frequency, Mason et al. (2009) measured flux densities at 31 GHz for 3165 NVSS sources using the
Green Bank 100m and the Owens Valley 40m telescopes. From these measurements they projected 
an integral source count at 31~GHz in the range 1 to 4~mJy, and found that the surface densities
implied account for $21 \pm 7$ per cent of the amplitude of the power detected in excess of intrinsic 
anisotropy by the Cosmic Background Imager (CBI) at $\ell > 2000$. We return to these 
results in \S~\ref{scest} and \S~\ref{disc}.

\begin{figure*}
\vspace{9cm}

\includegraphics{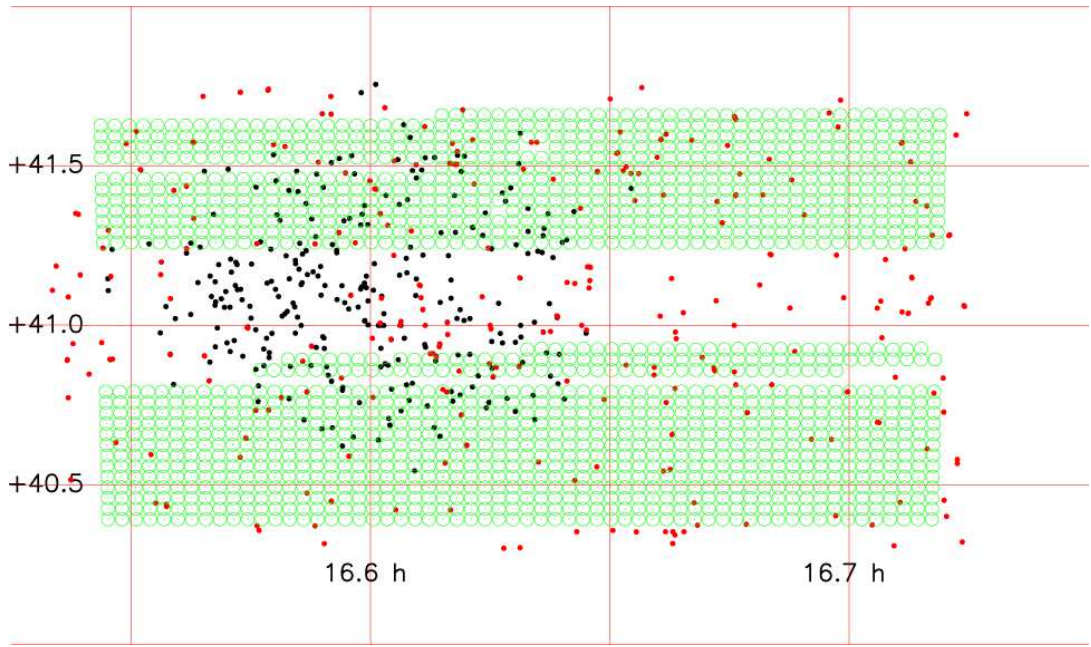}

\caption{The observing grid (Dec vs. RA, J2000) for our 1627 independent 
snapshot observations with the VLA at 43 GHz. The green circles represent the approximate 
size of the synthesized fields; the field-centre separation is twice the 
FWHM of the primary beam. 
Red dots mark source positions from the FIRST catalogue (Becker et al. 1995); black dots show 
source positions from the ELAIS N2 1.4-GHz catalogue (Ciliegi et al. 1999).}

\label{obsgrid}

\end{figure*}

%The paper describes design of the VLA 43-GHz experiment, observations and analysis, finding 
%sources, sorting real sources from spurious responses, and a methodology for extracting 
%count information in the face of multiple spurious responses. It concludes with a summary
%diagram of the 43-GHz source count plus brief discussion.

%\section{Road Map}

%\label{roadmap}

Surveying at frequencies as high as 43~GHz is not straightforward, as this pilot study
will demonstrate. Hence this paper does not follow the  conventional 
survey, analysis, catalogue, source-count etc. route --  the data do not lend themselves to 
such a procedure. We thus present a road map for orientation.
\begin{enumerate}
\item \S~\ref{surdes} {\em Survey Design} considers why we chose the region ELAIS N2 and  
how we came to the plan for coverage in terms of independent snapshots of given cycle parameters. 
The Appendix A1 describes an analysis leading to this design.
\item {\em Observations and Reductions} (\S~\ref{obsred}) constitutes the conventional 
part of the paper. We used standard AIPS analysis routines.
\item The 1627 independent snapshot images show many source-like deflections, as decribed in 
\S~\ref{obsred}. With our lack of knowledge of the sky at 43~GHz, we try in \S~\ref{dets}, 
{\em Finding Real Sources} to use as many avenues as possible to explore the reality of these
deflections. To do so, we select a sample of deflections from all snapshot images down to a 
level  of 3 times the rms noise level on each image. It is essential to emphasize at the outset 
that {\em we never use a 3$\sigma$ sample in a source-counting process}, as the effects of 
Eddington bias are lethal at such levels. However the sample gives us the broadest scope
to establish the frequency of real detections, and we try to do this by examining the
statistics of the sample, the coincidences of deflections with known sources from radio
catalogues at other frequencies, and the coincidences with optical counterparts using the 
Palomar II Sky Survey and the Sloan Digital Sky Survey.
\item The minimal returns from the exercises of \S~\ref{dets} led us to consider
how to use the survey data to estimate the source count at 43~GHz. We do so in \S~\ref{scest} 
by analyzing
the imprint of the primary beam shape on the expected radial distribution of sources in the
collective images, and we develop a technique for this that uses {\em each deflection},
real or otherwise, in a maximum-likelihood analysis. 
\item Finally (\S~\ref{disc}) we compare the results of this analysis with estimates of counts at other
mm-wavelength frequencies.
\end{enumerate}

\section{Survey design}
\label{surdes}

We chose the ELAIS N2 region as having a wealth of information available from surveys at radio to
X-ray wavelengths (see e.g. Rowan-Robinson et al. 2004). Moreover a bright calibrator (3C\,345, 1642+398)
stands close by, essential for observations at this frequency. 
 
To find the maximum number of sources, given source counts of slopes with which we are familiar in the
radio regime, the worst survey strategy is a single deep exposure, which we estimate would have 
yielded $\le$~0.05 sources in the primary beam (1.00 arcmin Full-Width Half-Maximum = FWHM) area, 
assuming a 24-h integration and an rms of 250~$\mu$Jy in 10~min. At radio wavelengths, wide and 
shallow always beats narrow and deep;  an analysis in Appendix~A1 quantifies this and shows that our 
adoption of the NVSS observing cycle (23~sec integration, 7~sec telescope settle) is near optimal 
for a series of independent snapshots. We estimated from the Ryle Telescope 15-GHz source count 
available at the time that observing 2880 fields in 24h should yield about 5~sources at a 4~mJy 
($\sim4\sigma$) survey limit. There is no significant difficulty in analysis of such a number of 
fields; this is essentially a mapping process and the general emptiness of the sky at 43~GHz 
implies that almost no deconvolution is required.

We used the technique of `referenced pointing' (see www.vla.nrao.edu/memos/test/189/) 
to position the primary beam with an accuracy of better than 5 arcsec.

\section{Observations and Reductions}

\label{obsred}

The observations were made in late 2001, at which time all VLA antennas had been newly equipped with
43-GHz receiving systems. In order to maximize sensitivity to extended emission and to 
minimize the effects of the atmosphere on phase stability, we used the the smallest VLA, i.e. 
the D configuration; and we observed in November, as the necessary phase stability at this frequency 
is generally available only in winter. We observed in 4 6-hour runs to keep to higher elevations,
so as to minimize atmospheric effects.

During the 24 hours of observing allocated to the programme, we made 1627 
independent snapshot observations, on the grid shown in Fig.~\ref{obsgrid}. 
Field centres are separated by 120 arcsec, twice the primary-beam FWHM. The
observations were carried out 02, 03, 12 and 23 Nov, 2001, in
outstanding weather conditions for the first three sessions, clear or
less than 20 per cent cloud. The final 8-hour session was disrupted by rain,
snow and cloud, with wind forcing the array to be stowed for much of
this period. This resulted in the gap at the central declinations of Fig.~\ref{obsgrid}. 
About 19.0 useable hours of observations were obtained 
as judged from the phase stability, and out of 24h this is good
fortune at frequencies as high as 43~GHz.

The primary beam FWHM is 1.00 arcmin and the synthesized beam 
FWHM 2.3 arcsec. The phase calibrator 3C345 (1642+398) was observed approximately every 30 minutes.
For flux density calibration, once or twice per session we observed 
3C\,286 (1331+305), with a flux density of 1.45~Jy at 43.4 GHz. The effects of slight resolution 
at this frequency were removed via the appropriate VLA model supplied in the NRAO AIPS software
package.

We used standard AIPS procedures in the analysis. At each field centre we made a
map of $1024 \times 1024$ pixels each 0.2 arcsec square, total map extent $3.4 \times
3.4$ arcmin. We used a minimal CLEAN (100 cycles, gain=0.1, $\le$80 components 
$\ge$1~mJy) on each map; this procedure unfortunately proved 
necessary in order to use subsequent AIPS routines, which require CLEANed data as input. 
%Adding this possible non-linearity was not a desired feature of the analysis.

In order to examine whether our phase calibration
every $~$30 minutes was adequate, and in particular to look at potential decorrelation,
we carried out a simple experiment on the phase 
calibration data of the observing day 12 Nov 2001. The phase calibrator was observed for
one minute at UT times of 17:11, 17:28, 17:47, 18:04, 18:27 and 18:43. Using AIPS and 
doctoring the input files, we made three measures of
the calibrator flux density, assuming it was a programme source and using phase calibrators either
side of it for flux calibration. We thus made a measure of the flux of 1642+398 at 17:28
using the 17:11 and 17:47 observations alone as flux calibrators, and so on for meaures
of 1642+398 at time of 18:04 and 18:27. The three values of flux density obtained
differed from the input value (9.11 Jy at 43.4 GHz) by +1.5\%, +0.1\% and +2.7\%. We 
concluded that calibration was satisfactory; these differences are insignificant in 
comparison with the noise and resolution effects measured for the source-like deflections 
encountered in the survey.

The final 1627 maps have a median rms of 1.4 mJy, close to that estimated for the survey. 
The lowest noise maps have an rms of 0.9 mJy; 67 per cent have rms noise $\le 1.5$~mJy and 
98 per cent $\le 2.0$~mJy.
Inspection of the maps shows a number of low-level striations 
well under the rms noise level, and responses, many of 
which look source-like. The responses happen because the receiver or sky noise is only introduced 
at points with measurements in the $uv$ plane. For snapshot observations in
particular, this means that with noise partially correlated between antennas 
the images will show apparently coherent
structure even in the absence of any real sources.  This will be in the
form of intersecting ripples of various periods (one per $uv$ point) with
amplitude and phase fluctuating between observations. Responses, fictitious sources,
are most easily generated where these ripples intersect.  If all source-like responses are
spurious and  due to 
receiver noise as described, then there is no imprint of the primary beam in the image 
from each snapshot, a point discussed in detail in \S~\ref{stats} and \S~\ref{scest}.  Most of
these responses are very extended to the 2.3-arcsec beam but many look like unresolved
sources. We consider this issue in the following section.

In short, the autocorrelation function of the images is not remotely Gaussian, although
the rms noise (in good weather conditions) is close to that predicted from telescope and 
instrumentation parameters.

%There was good physical reason to confine our attention to the compact `objects'. All
%extended-emission
%radio sources are known to have synchrotron $\sim$ power-law spectra of spectral index $\alpha$
%($S \propto \nu^{-\alpha}$) with $\alpha < -0.5$, so that any sources detected at 43 GHz will almost
%certainly not show extended emission or be associated with extended radio objects detected at lower
%frequencies. Moreover many if not most such objects will be resolved out by the $\sim$2-arcsec synthesized
%beam of the survey. We are looking for compact objects, QSOs or BL\,Lac objects  with flat, 
%inverted and/or
%variable emission sources, probably beamed, and with structures on milliarcsec VLBI scales.

\begin{figure}
\vspace{7.cm}

\includegraphics{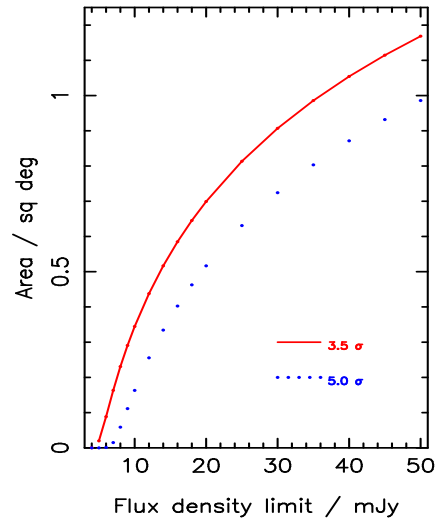}

\caption{The area surveyed down to a given sensitivity level, where this is taken to be
either 3.5 or 5 times the rms noise on each of the 1627 snapshot maps, and the primary beam is
taken to be a circular Gaussian with FWHM 60.0 arcsec. It is known that out to a radius of 1~FWHM, 
this assumption represents the beam well.}

\label{area_mJy}

\end{figure}

The total area covered as a function of flux density was
calculated by considering each snapshot field separately, the primary beam and the rms flux
density, and adding up the areas available to detect flux density for point sources above given 
signal-to-noise ratio (s/n). The resulting curves of area versus sensitivity are shown in 
Figure~\ref{area_mJy}.

\section{Finding real sources}

\label{dets}

We applied the AIPS routine SAD to each map to search for sources, setting a 
flux limit of 3.0 mJy. We then removed from the subsequent list all objects with peak 
flux-density to noise ratio below 3.0. This gave us a uniform set of 54225 
deflections (an average of 33 per field). Of course a 3$\sigma$ limit is far 
too low if we were selecting responses
for a source list based on s/n alone. Under such circumstances, Eddington bias is extremely damaging
and even a 4$\sigma$ limit results in significant contamination of a source list with noise peaks. 
However, here we are simply selecting a sample to search for possible detections using criteria other 
than s/n. 

The net result of the extensive searches described below is one secure 43-GHz source, the 
previously catalogued QSO 16382+414 with S$_{43\,{\rm GHz}}$ = 26~mJy, and 
several possible 43-GHz sources coinciding with either blue stellar objects or sources previously 
catalogued at 1.4~GHz.

%The number of such $3.0\sigma$ detections per field is shown in Figure~\ref{sigs}.

We reasoned towards establishing real sources detected in the survey 
via the following steps. 

\subsection{Source statistics} %section 4.1

\begin{figure}
\vspace{6.5cm}

\includegraphics{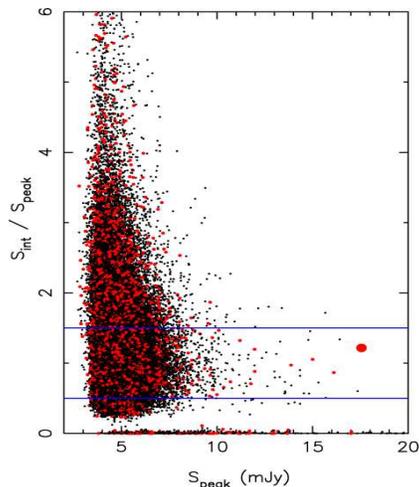}

\caption{The $3.0\sigma$ `detections' on the 1627 maps, shown in terms of ratio of
integrated to peak flux density vs. peak flux density. Red dots are `detections' 
within a radius of 60~arcsec = 1~FWHM of the field centres; black dots are `detections' at larger 
distances. Real sources are expected lie within the FWHM distance, and to be unresolved, lying 
between the horizontal lines at ratios of 0.5 and 1.5. The points along the horizontal
axis represent single-pixel responses which are rejected from further consideration by the lower 
ratio limit. The large red dot represents 16382+414, 
the known QSO detected in the 43-GHz survey; see \S~\ref{coinc}.}

\label{rat}

\end{figure}

\label{stats}
Fig.~\ref{rat} shows a diagram for the 54225
`sources' with ratio of integrated to peak emission (`Size Factor' or SF) as a function of peak flux 
density. The diagram demonstrates that attention should be confined to those objects for which this 
flux ratio is less than 2, and that there may be a set of unresolved
detections at $S_{\rm peak} > 7$~mJy, $0.5 < $ SF $ < 1.5$ whose reality needs to be 
examined further. 

Responses narrower than the synthesized beam, SF $<$ 0.5, are probably spurious; a number are clearly 
single-pixel fluctuations. We examined a fair sample of the `detections' both extended and apparently 
point-like, via visual inspection and profiling of a sample of $3.0\sigma$ detections on the synthesized maps. Most do not look convincing, sitting astride response 
lines or at the junctions of these lines.  There is good reason to confine our attention to the compact `objects'. All extended-emission radio sources are known to have synchrotron $\sim$ power-law spectra
of spectral index $\alpha < -0.5$, so that sources detected at 43 GHz will almost
certainly not show extended emission or be associated with extended radio objects detected at lower
frequencies. Moreover many if not most such extended objects would be resolved out by the
2.3-arcsec synthesized beam of the survey. We are looking for compact objects, QSOs or BL\,Lac
objects with flat, inverted and/or variable spectra, probably beamed, and with structures on 
milliarcsec VLBI scales. However in the interests of objectivity the first of our investigations here 
retains the entire sample of 54225 responses.

%\special{psfile=dets_per_fld.eps hoffset=5 voffset=180 vscale=33 hscale=33
%angle=-90}

%\caption{The histogram of numbers of $3.0\sigma$ detections on each of the 1627 maps. A 
%Poisson distribution with $\mu=12.5$, approximately the median number per field, provides 
%a reasonable fit.}

%\label{sigs}

%\end{figure}

\begin{figure}
\vspace{6.5cm}

\includegraphics{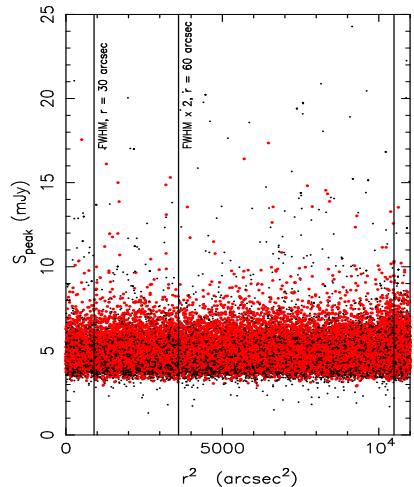}

\caption{The $3.0\sigma$ `detections' on the 1627 maps, plotted against the square of the distance 
in arcsec from each field centre. Red dots are detections for which the ratio of 
integrated to peak
flux lies between 0.5 and 1.5, i.e. unresolved to the 2.3-arcsec synthesized beam; black dots are all 
other deflections. There is no noticeable concentration of the high s/n deflections to the primary 
beam response denoted by vertical lines corresponding to the FWHM and $2\times$FWHM; the distribution 
looks uniform with $r^2$, indicating that most responses do not represent real sources.}

\label{speak}

\end{figure}

Figure~\ref{speak} argues strongly against the reality of most of these `sources', even the stronger
ones. Here deflection height is plotted against $r^2$, the square of the distance from the individual
field centres in arcsec. This is effectively area, so that the plot should be uniform if the
deflections are random results, e.g. of intersecting response lines. The vertical lines show the value of 
$r^2$ corresponding to the primary-beam half-power point, $r^2$ corresponding to twice this,
and $r^2$ corresponding to half the width of the synthesized square. Deflections corresponding
to real sources should be concentrated within $r^2_{\rm FWHM}$, and few if any should 
appear beyond $r^2_{\rm 2FWHM}$. The deflections, compact or extended, appear to be depressingly
uniform across the full extent of $r^2$. We must conclude that the vast majority
cannot represent real sources.

\subsection{Coincidences with catalogued sources?} %section 4.2
\label{coinc}
We compared all 54225 `source' positions with those for radio sources in the ELAIS~N2 
catalogue (Rowan-Robinson et al. 2004) from the 1.4-GHz deep surveys of this region 
(Ciliegi et al. 1999; Ivison et al. 2002; Biggs \& Ivison 2006), 
and with the sources of the FIRST survey (Becker et al. 1995) There were a total of 538 such sources 
within our synthesized area (Figure~\ref{obsgrid}).  
We confined attention to those within 10 arcsec of
catalogued sources; the properties of these objects were found not to differ from the 
others. We then examined carefully all the objects lying within 2 arcsec of the catalogued
sources. The total coincidence area corresponding to 2 arcsec radius suggests a possible 5.4 
coincidences due to chance. We found a total of 15 such coincidences. We then confined attention 
to objects with size factors of 1.5 or less; the list of such coincidences totalled six.

(With regard to this SF $< 1.5$ upper limit, it could be argued that it should be relaxed to 
somewhat larger values at lower signal-to-noise. We carried out Monte-Carlo trials, and found that 
although as expected for point sources there is a distinct increase in observed SF for true point 
sources as signal-to-noise is lowered to the $3\sigma$ threshold, only some 5 per cent of the trials 
yielded SF values above 1.5. Thus the above list of 6 possible coincidences above may be short by 
half-a-source statistically; but on the other hand, relaxing the criterion feeds in a far greater 
proportion of assuredly spurious `detections'.)

\begin{figure*}
\vspace{6.0cm}

\includegraphics{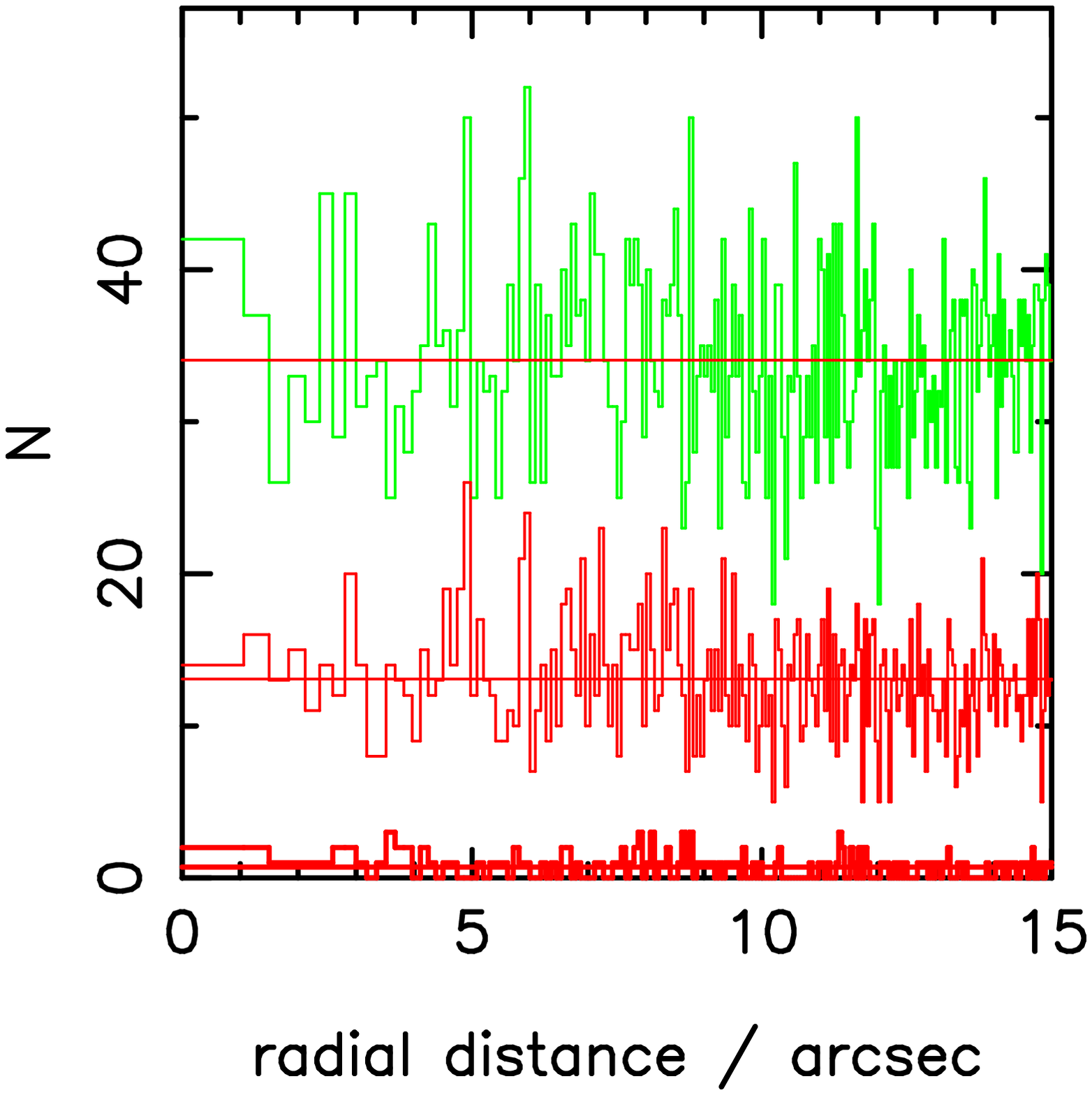}
\includegraphics{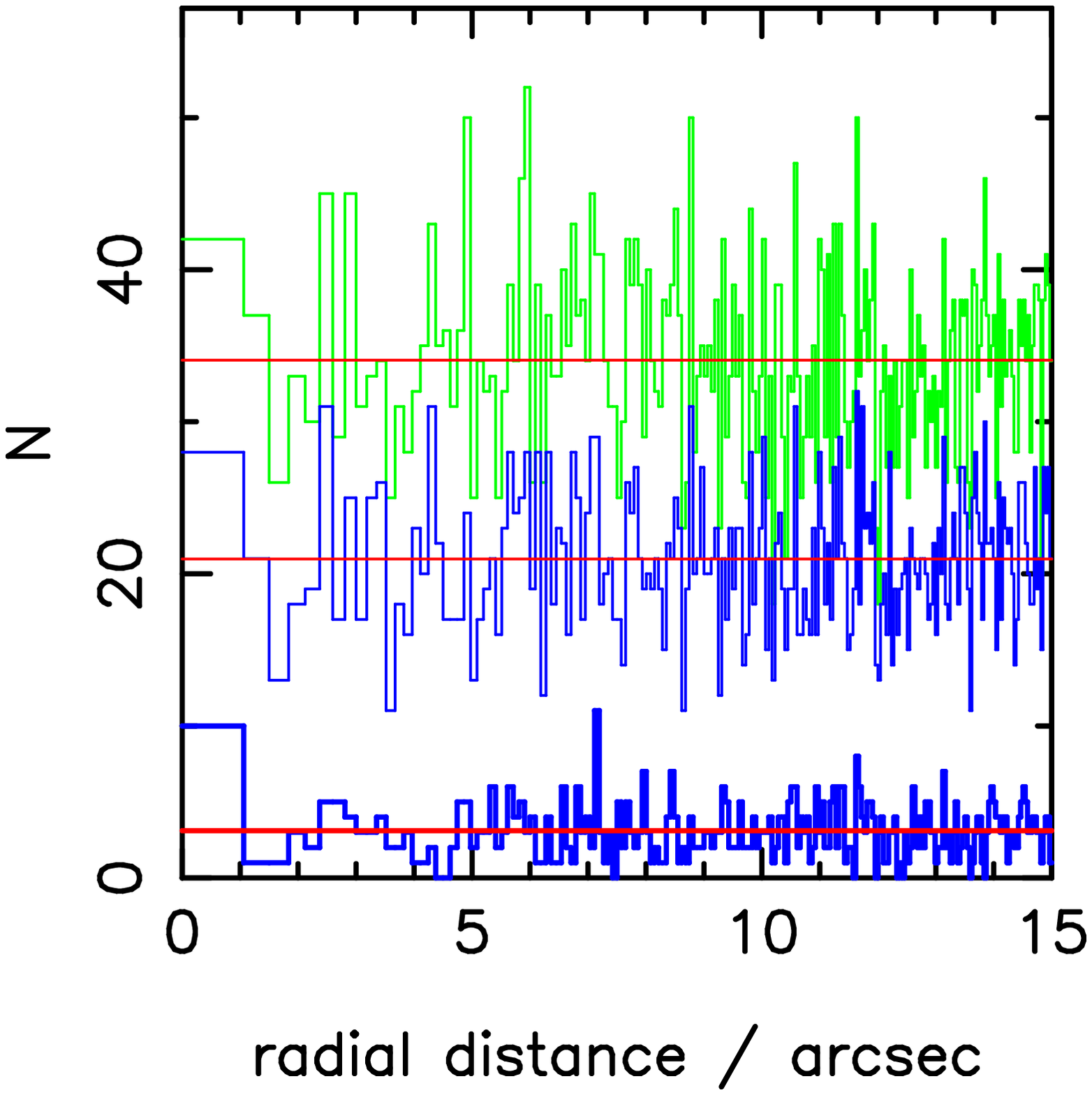}

\caption{The radial distributions for optical objects from the SSS in equal-area annuli about
the field centres for the sample of
16331 source-like deflections. Each annulus has the same
area as the central circle of radius 1.06 arcsec. Random association
of optical and radio objects should show a flat histogram with Poisson noise superposed. In 
both left and right panels the upper histogram (green) is for all objects down to the limit of 
the SSS survey, for source-like deflections within 80 arcsec of the synthesized field centres. 
Left panel: galaxies, as classified in the SSS object lists, are shown as the middle and lowest
distributions, the middle distribution for all galaxies, the lower for blue galaxies with 
$m_b - m_{r1} \leq 1.0$. Right panel: the central histogram is of all objects classified by 
SSS as stellar, while the lower histogram is for bluish stellar objects with $m_b - m_{r1} 
\leq 1.40$ and $m_{r1} \leq 20.2$).}

\label{rad_dist_ids}

\end{figure*}

Of these 6 coincidences within 2 arcsec, we rejected two as possibilities because the FIRST 1.4-GHz
contour maps suggest that the catalogued source is a component of a double-structure radio 
source. An inverted radio spectrum object has never been seen as a lobe of a double radio source
and it is thus highly unlikely that the 43-GHz deflection is real. Of the remaining four sources,
three lie more than 60 arcsec from their field centres, and none has a candidate optical identification 
down to the limits of the SDSS (Sloan Digital Sky Survey). 
The probability of finding real sources beyond one FWHM from the beam centre is minimal, and in each 
case, supposing the source is real, the beam correction would result in flux densities at 43 GHz of 
from 120\,mJy to $>$1\,Jy. Such extreme spectrum objects would have been detected should they exist; and 
the lack of any optical counterpart above SDSS limits results in a further drastic reduction in 
probablity that these objects are anything other than artifacts. The closest of these to the field
centre, 16392+408, lies 62 arcsec off axis, and correction for the primary beam attenuation
would give it a flux density at 43.4 GHz of $\sim$120~mJy. The coincident FIRST source has a flux
density of 22.1~mJy at 1.4~GHz. It is entered in Table~\ref{coinc_all} as a candidate source, albeit 
an improbable one.

Of the 6, this leaves one final coincidence to consider, and here there is no doubt of the reality of the 
source at 43.4 GHz; it has a s/n of 9.6. The object is 16382+414, a flat-spectrum QSO known to have 
a redshift of 2.2, apparently variable as the Ciliegi et al. (1999) and FIRST flux densities 
(24.3 and 37.3~mJy respectively at 1.4 GHz) differ significantly. The details of this
object are in Table~\ref{coinc_all} (bold typeface). This real source is the strongest 
compact response within $0.5 \times$~FWHM = 30~arcsec of the individual field centres. It may be 
slightly extended to the 2.3-arcsec sysnthesized beam.

The yield (of $\sim$one coincidence) is small but instructive.
The secure coincidence with 16382+414 guarantees that the survey and reduction procedure is 
capable of finding real sources. It also ensures that the coordinate systems are in coicidence to
better than 1 arcsec, so
that other coincidences, either with previously catalogued radio sources or with optical
identifications, are possible. It does not rule out the possibility that some `sources' close
to the $3.0\sigma$ limit in s/n are real, but the coincidence results plus Figures~\ref{rat} 
and~\ref{speak} indicate that the great majority are not. 

\subsection{Optical identifications?}

The AT20G survey (Ricci et al. 2004; Sadler et al. 2006; Massardi et al. 2008) suggests that compact 
objects over the range of flux densities sampled by the survey (20 - 500 mJy at 20~GHz) are 
identified predominantly (50 to 80 percent) with blue stellar objects, QSOs or BL\,Lacs. 
A small proportion is identified with relatively bright 
galaxies. A further way then of searching for real sources amongst the
54225 deflections is to look for optical coincidences, primarily stellar counterparts. 
It is possible to do this in 
bulk because of the ease with which automated comparisons can be done with both POSSII and UKSTU sky 
surveys, thanks to the SuperCosmos Sky Survey (SSS).

%\begin{figure}
%\vspace{9.0cm}

%\special{psfile=qband_quh_comp.ps hoffset=-10 voffset=-15 vscale=40 hscale=40 angle=0}

%\caption{A plot of $m_{R2}$ vs $(m_B - m_{R2})$ for the QSOs from the 2.7-GHz sample of \citet{jac02}. The
%dots plot the magnitude and colour data for these Flat-Spectrum Radio QSOs (FSRQs) obtained from the 
%SSS exactly as for the optical objects in the fields of the 16331 43-GHz survey point-like deflections. 
%The stellar objects within 1.06 arcsec (the central circle) of the 43-GHz deflections are shown as open 
%circles. The star symbol indicates the one certain QSO found in the 43-GHz survey. The region above 
%and to the left of the horizontal and vertical lines suggests selection criteria for the 43-GHz 
%objects within the 1.06-arcsec radius which most resemble the distribution found for the Jackson 
%et al. FSRQ sample.}

%\label{quh_comp}

%\end{figure}

To carry this out, we reduced the sample of 54225 as follows. We first placed a stringent SF criterion
to confine ourselves to compact `detections', using only those `sources'with $0.6 < {\rm SF} < 1.4$.
This reduced the sample to 26090. (Again it could be argued that we are throwing away possible 
real sources because of the increasing spread in SF at lower signal-to-noise; again we argue that we are 
after a set of unambiguous detections, which if present would lead us to carry out a more rigorous source 
cataloguing process. We found no encouragement to do this.) We then threw away responses in the  
field `corners' by setting a generous radius limit of 80 arcsec, 
(c.f. half-power beamwidth of 60 arcsec); deflections further off centre than this require 
(poorly known) beam-factor corrections so large as to make them Jy-level sources. 
The sample is now 16331 in size. We examined all 16331 position on SSS R-band 
images, easy to automated with the provision of object catalogues in the SSS download which list 
magnitudes, positions, and object type (galaxy/stellar discrimination) for all objects above the
detection limit in each postage-stamp image.

%\begin{figure}
%\vspace{6.cm}

%\special{psfile=src-list.ps hoffset=-10 voffset=-15 vscale=40 hscale=40
%angle=0}

%\caption{The area mimics the size of each synthesized field at 43 GHz. The objects plotted are the stellar
%objects as selected a) from proximity within 1.06 arcsec to the positions of 43-GHz point-like deflections, 
%and b) according to the colour -- magnitude criteria of Figure~\ref{quh_comp}. Red dots have 
%$1.0 < B - R_2 < 1.4$, and blue dots $B - R_2 < 1.0$. The star symbol represents the 
%single known QSO of the 43-GHz survey.}

%\label{coinc_dist}

%\end{figure}

Figure~\ref{rad_dist_ids} shows a histogram of surface densities of the optical objects 
in equal-area annuli, 200 annuli out to a 
distance of 15~arcsec from each deflection position. The beam is small enough (FWHM 2.3~arcsec) so that 
true identifications should lie in the first area, the central circle of radius 1.06 arcsec. 
The surface distribution of all optical object types (upper histograms) shows a positive signal 
in this bin of little significance. If the distribution of objects classed in SSS as galaxies 
is considered (left panel) there is nothing of note -- apparently no coincidences (although 
statistically a few identifications, say $\leq 5$ cannot be ruled out). The right panel tells a 
different story for stellar objects, which we might expect to be QSO or BL\,Lac identifications -- the 
distribution  for all `stars' shows an excess in the central bin at a level of $2.8\sigma$.
We examined the distribution of SSS $m_b - m_{r1}$ colours vs $m_b$ for the sample of FSRQ of 
Jackson et al. (2002) and this shows that FSRQ are confined to a region of the colour -- magnitude plane 
with $(m_b - m_{r1}) < 1.4$, $m_{r1} > 20.2$. (We used the SSS magnitude system, 
in which our $m_b = B_j$, and our $r_1 = ESO-R/POSS-I E$.) Applying this cut to the stellar 
objects (and in addition requiring that the objects lie within 60 arcsec of the field centre) yields the 
lower distribution in Figure~\ref{rad_dist_ids}, right panel. The
excess in the central circle now runs at $4\sigma$; there are 10 objects and the mean value is 3.
The implication is that a few ($\leq 7$) of the the blue stellar objects may be real 
identifications and that this number of deflections out of the 16331 may be real sources. The QSO
16382+414 is amongst these coincidences, rediscovered yet again by this automated identification 
technique, and as well, one of the 10 objects proved to be a galaxy when examined with the higher
resolution of the SDSS.

%Figure~\ref{coinc_dist} shows where these stellar objects lie in relation to the centres of the synthesized
%fields. It might have been hoped that this distribution would be much more centrally concentrated, and that
%the selection process above would have resulted in a distribution resembling that of the left panel in 
%Figure~\ref{sim}. In fact the closest object to the field centre is the confirmed source, the known QSO.
%However, there is a signifcant trend of the bluer objects to be closer to the field centres than the
%redder objects. When all stellar coincidences are considered, ie no colour --  magnitude selection as in
%Figure~\ref{quh_comp} is made, this trend is statistically significant. Thus both on the basis of the
%coincidence statistics (Figure~\ref{rad_dist_ids}) and the radial distribution (Figure~\ref{coinc_dist}) 
%there is clear indication that a very few, perhaps 3 or 4, of the deflections are real sources with blazar
%counterparts. 

\begin{figure}
\vspace{7.5cm}

\includegraphics{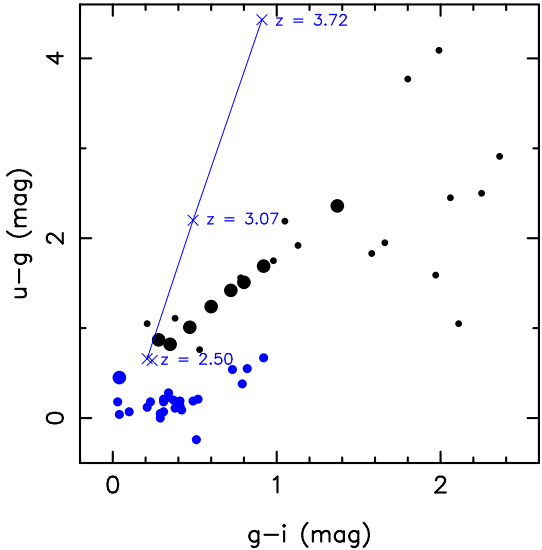}

\caption{Colour -- colour diagram using SDSS photometry for 27 of the 28 objects which 
are stellar and within 1.06 arcsec of deflection positions. (SDSS shows one of the 28
`stellar' objects from the UKST SSS to be a galaxy.) Of these objects 26/27 are shown as black 
filled circles; the 8 larger of these are the objects selected from the SSS catalogues as having 
$m_b - m_{r1} \leq 1.4$ and $m_{r1} > 20.2$. The known QSO identification in the set of 27 
stellar objects (16382+414) is shown as the large blue filled circle. Smaller blue filled circles
represent the SDSS photometric data for 26 FSRQ from the sample of Jackson et al. (2002) which
lie in the area covered by SDSS. These cluster at values of $u-g$ significantly
below the star relation, with four exceptions: the four higher-redshift ($z > 2.5$) QSOs
marked as crosses shoot to high values of $u-g$ because of the increasingly large $u$ 
magnitudes due to absorption below the Lyman-$\alpha$ line (Weinstein et al. 2004).}

\label{sdss_cols}

\end{figure}

Using SDSS photometric data, we plotted the remaining 8 stellar objects in the central annulus (10, minus
the QSO 16382+414, minus the galaxy) in Figure~\ref{sdss_cols} as the 
large black filled circles. The known QSO is plotted as the large blue 
filled circle. In the same plot we also placed the photometric colour data (black dots) for the 27 of 
the 28 objects 
falling within the inner circle of 1.06 arcsec radius (central distribution in right panel of 
Figure~\ref{rad_dist_ids}), again omitting the galaxy; the 9 objects (8 stellar, one QSO)
already plotted are of course a subset of these 28 objects. These objects, particularly the 
brighter ones, generally 
follow the black-body locus for hot stars. As a control sample we plotted the SDSS $(u-g)$ vs. $(g-i)$ 
colours for a sub-sample 
of FSRQ from the Jackson et al. (2002) catalogue, namely those from the full sample which fell 
within the SDSS area, a total of 26. These generally cluster at values of $(u-g)$ {\it below} 
the stellar sequence, and the known 
QSO in the sample of 27 stellar objects falls in with these other FSRQ. The few FSRQ of the 
Jackson et al. (2002) sample with redshifts greater than 2.5  (crosses) streak away to high values of 
$u-g$ as is well 
known (see Weinstein et al. 2004); the absorption shortward of Lyman-$\alpha$ gives relatively larger 
values of $u$). 
There is a small region near $(u-g) = 1.0$, $(g-i) = 0.4$ which appears common to the stellar sequence 
and to FSRQ with $z \sim 2.5$. Some 6 of the stellar objects coincident with 43-GHz deflections lie 
close to this region, and might be considered as possible identifications; this number would certainly 
account for the bulk of the statistical excess. Three of these are in the original sample of 8 likely
candidates, while of the other three, two have distances from the filed centres of just over 60 arcsec and 
the third is fainter than the $m_{r1} = 20.2$ limit. 

There are, however, indications that most of these objects are not identifications and the deflections are 
consequently not real sources. 
\begin{enumerate}
\item Stacking the FIRST FITS images of these 6 objects, or the 8 objects of
the most probable identifications, or indeed the 26 stellar objects within the 1.06 arcsec radius, yields
total absence of any central response. Mean flux densities at 1.4 GHz must 
lie below levels of $\sim 0.1$~mJy, leading to putative spectral indices of $\alpha_{1.4}^{43} > 1.5$. 

\item If the deflections were real sources, then they should cluster within the FWHM
of the primary beam, i.e. within 30 arcsec of their field centres, and certainly within a radius of 2FWHM
(as in Figure~\ref{sim}, left panel). 
In fact the radial distribution of the 26 objects relative to their field centres is as follows: 
within 20 arcsec -- 2, between 20 and 40 arcsec -- 4, between 40 and 60 arcsec -- 10; between 60 and 80 
arcsec -- 10. The corresponding numbers for random distribution, i.e. according to annular area, 
are 2, 5, 8 and 11. The distribution thus appears random; there is no indication that these `sources' 
were selected by beam response. This is equally true for the sub-sample of 8 likely identifications, and
the sub-sample of 6 objects with colours close to the FSRQ locus. 

\item Finally, all the redshifts would need to be clustered in a small range of z, i.e. $2.5 \pm 0.2$. 
This is unlikely; the range is not at the peak or centroid of the redshift distribution for 
FSRQ (Jackson et al. 2002).

Despite the low probability that these 6 `sources' are real, they have been listed in 
Table~\ref{coinc_all}.

\end{enumerate}

We have to conclude that the majority and possibly all of these candidate identifications (exept for the 
known QSO) are chance 
coincidences, with the high response in the central bin of Figure~\ref{rad_dist_ids} (right panel) 
a statistical anomaly. It could still be that in addition to the known QSO, one to three of 6 
sources listed as stellar in Table~\ref{coinc_all} may be real. Direct observations, either 
confirmatory radio observations or optical spectra of the putative counterparts, are
needed in this respect. Finding charts are readily available using the SDSS 
navigate tool http://cas.sdss.org/astrodr7/en/tools/chart/navi.asp.

\begin{table*}

\begin{center}

\caption{Source list: coincidences with stellar counterparts and catalogued sources}

\begin{tabular}{cccrcccccccc}
%source ra dec dist speak scorr s1.4 alpha g u-g g-i type  
\hline\hline
Source&RA & Dec &dist$^1$&$S_p(43)^2$&$S_{\rm corr}$(43)$^3$&$S$(1.4)$^4$&$\alpha^5$&$g^6$&$u-g^6$&$g-i^6$&type$^7$\\
\hline
        &&&&&&&      &       &            &           &             \\
%16:36:19.73 & -0.42 & 40:22:55.0 & -1.70 & 1.75 &  9.8 &  68.5 & $>$1.23 & 19.93 & 0.34 & galaxy \\
%16:38:11.71 &  1.15 & 41:28:41.8 & -0.89 & 1.45 &  7.3 & 638   & $>$2.09 & 19.56 & 0.67 & stellar\\
%16:38:17.32 &  0.91 & 41:27:29.5 &  0.20 & 0.93 & 17.6 &  25.9 &   +0.02 & 19.55 & 0.41 & QSO, z=2.2\\
%16:38:20.87 &  0.93 & 41:38:34.6 &  0.85 & 1.26 &  7.3 & 137.5 & $>$1.64 & 19.64 & 1.14 & galaxy \\
%16:38:49.71 &  0.32 & 40:47:37.8 & -0.51 & 0.60 &  6.6 &   7.2 & $>$0.66 & 19.73 & 1.86 & galaxy \\

16338+404&16:33:49.92&40:25:21.9&51.5& 4.2$\pm$1.4&32.5&$<$0.1&$>$0.9&18.17&1.01&0.47 &stellar\\ %15821
%16345+404&16:34:30.96&-0.24&  40:25:37.4& 0.69& 0.73&39.5&  4.5&14.9&$>+0.67$& 1.23&  17.84& stellar\\
16345+406&16:34:32.18&40:41:47.6&59.5& 4.2$\pm$1.3&64.2&$<$0.1&$>$1.5&17.61&0.87&0.28& stellar\\ %10652
16345+415&16:34:34.06&41:30:55.8&39.3& 4.3$\pm$1.4&14.0&$<$0.1&$>$1.2&18.77&0.82&0.35& stellar\\ %14433
16349+405&16:34:58.73&40:32:34.3&59.7& 5.1$\pm$1.7&79.4&$<$0.1&$>$1.9&21.75&1.11&0.38& stellar\\ %13314
16361+405&16:36:06.44&40:31:24.0&53.2& 4.8$\pm$1.6&42.4&$<$0.1&$>$1.8&22.04&0.76&0.53& stellar\\%15203
16377+416&16:37:47.27&41:38:32.4&61.3& 4.2$\pm$1.4&75.9&$<$0.1&$>$1.9&19.57&1.05&0.21& stellar\\ %12592
%16358+404&16:35:49.14& 0.59&  40:25:15.1& 0.58& 0.83&33.9&  4.0& 9.8&$>+0.62$& 1.30&  20.03& stellar\\
%16363+407&16:36:19.40& 0.00&  40:45:13.6& 0.22& 0.22&41.8&  4.2&16.2&$>+1.21$& 1.00&  18.45& stellar\\
%16382+406&16:38:15.80& 0.09&  40:36:19.5&-0.55& 0.56&58.3&  5.1&70.2&$>+1.20$& 1.23&  19.21& stellar\\
{\bf 16382+414}&16:38:17.32&41:27:29.5&22.4&17.6$\pm$1.8&25.9&  37.3& -0.11&19.93&0.45&0.06& QSO, z=2.2\\
16392+408&16:39:12.05&40:52:35.8&61.9& 6.2$\pm$1.8&118.5& 22.1& +0.49& - & - & - & FIRST coinc\\
%16406+134&16:40:36.40&-0.39&  41:18:20.1& 0.62& 0.73&47.3&  6.2&34.4&$>+0.91$& 1.36&  18.32& stellar\\
%16413+404&16:41:20.18&-0.91&  40:24:16.6& 0.52& 1.04&54.9&  5.0&51.2&$>+1.03$& 1.10&  17.27& stellar\\

\hline

\end{tabular}

\label{coinc_all}

\end{center}
\begin{tabular}{l}
$^1$Arcsec distance of deflection from primary-beam axis.\\
$^2$mJy peak deflection flux density at 43.4~GHz.\\
$^3$mJy 43.4-GHz flux density corrected for primary beam attenuation.\\ 
$^4$mJy 1.4-GHz flux density from FIRST survey: catalogue or stacked images.\\
$^5$Spectral index 1.4 - 43.4 GHz, $S \propto \nu^{\alpha}$.\\
$^6$Magnitudes, from SDSS photometry.\\
$^7$stellar -- coincidence with blue stellar object to within 1.06 arcsec; QSO -- catalogued quasar; 
FIRST coinc -- coincidence with\\ 
\ \ catalogue source, FIRST survey.
\end{tabular}
\end{table*}

\section{A statistical estimate of the 43-GHz source count}
\label{scest}

There is a way to use the deflection statistics alone (randoms and reals) to estimate the 
source count. This is based on the primary beam response, which should leave a significant 
imprint on source distribution if even a small proportion of the sources are real, so small 
a proportion as not to be detectable in individual fields. The effect is dependent 
on source-count slope, enhanced when the source count is steep.

\begin{figure*}
\vspace{5.7cm}

\includegraphics{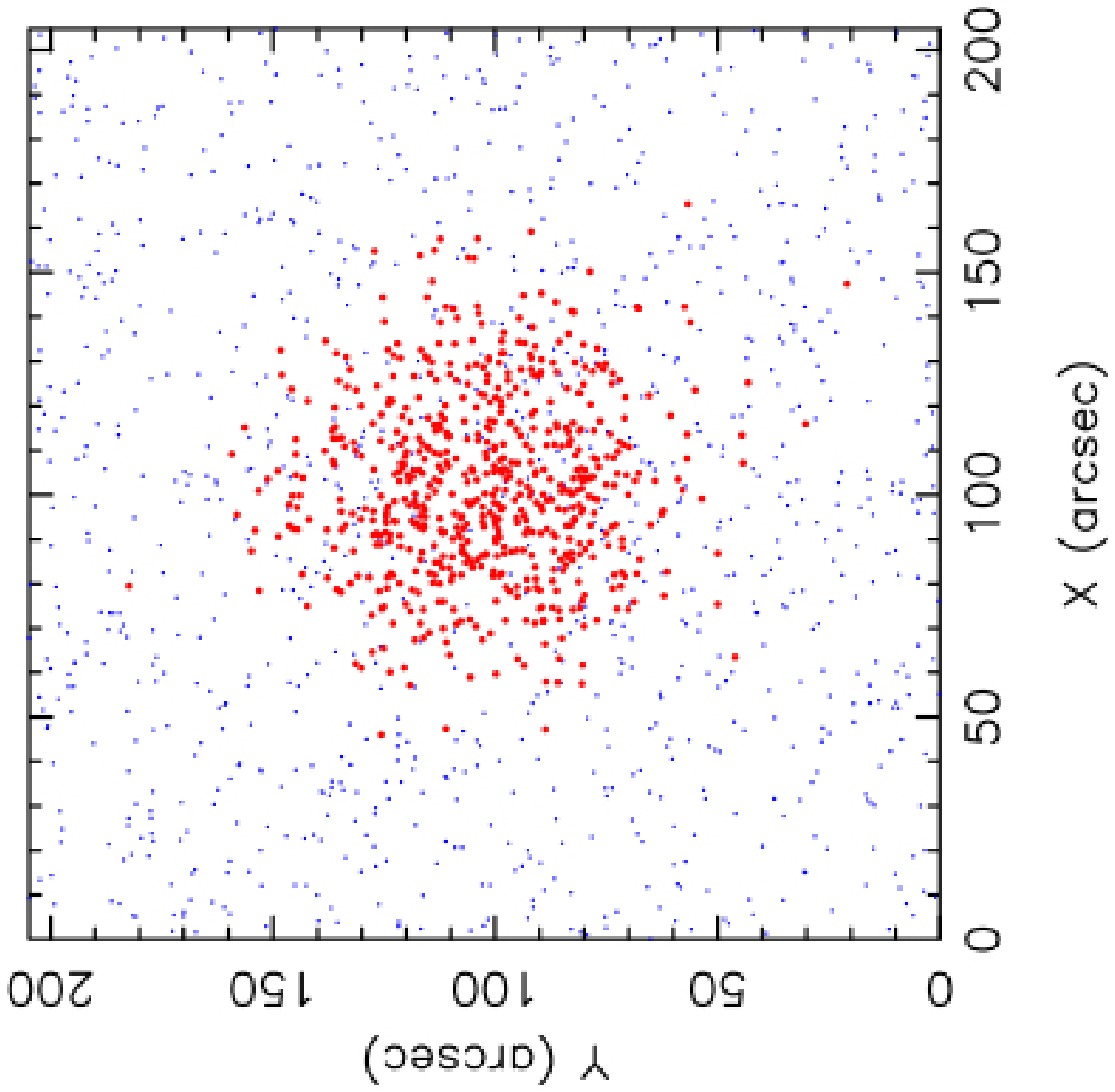}
\includegraphics{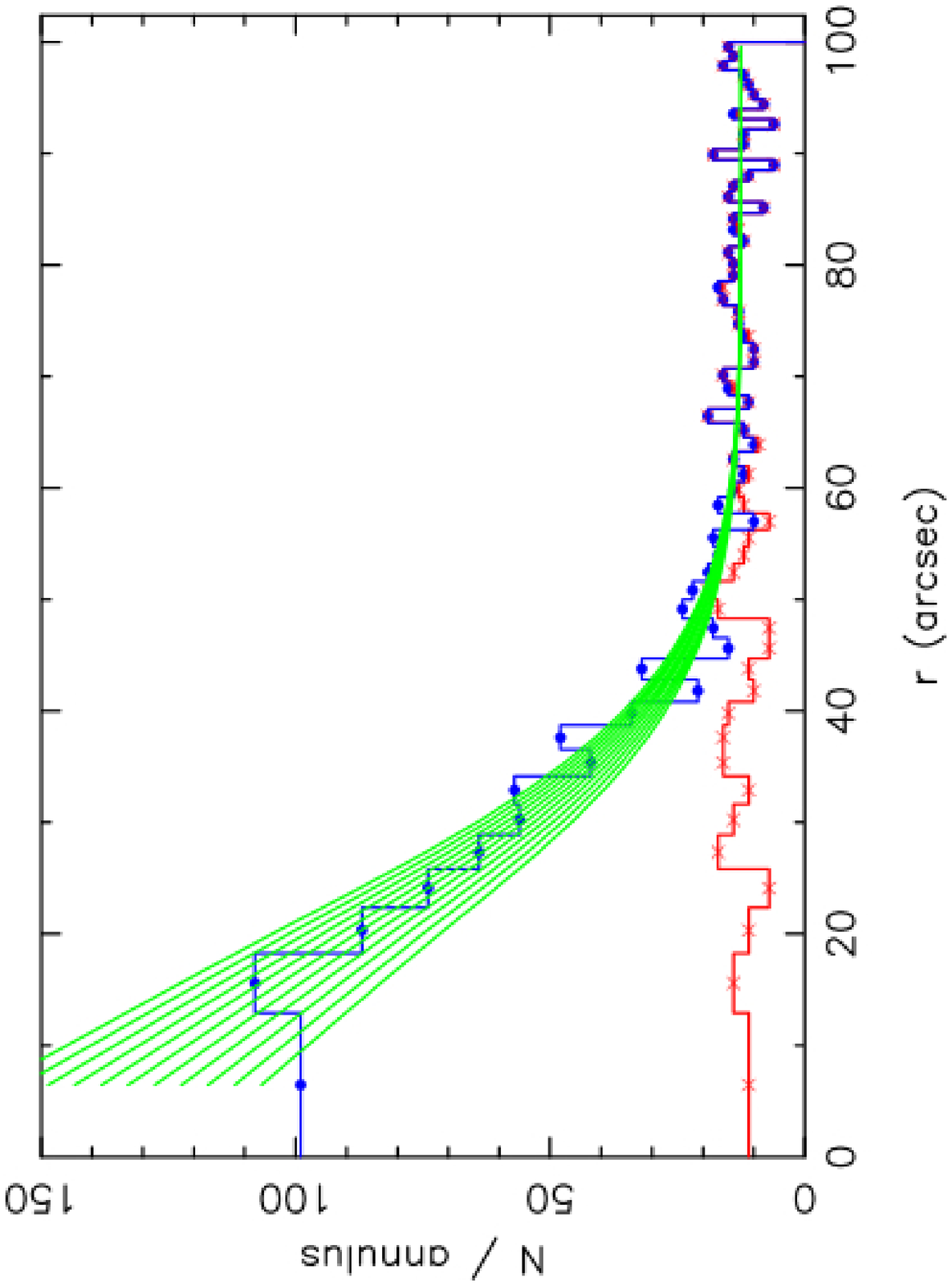}

\caption{A simulation to show how primary beam response can be used not only to distinguish the
proportion of true detections, but also to measure source count parameters. The simulated field
(to the scale of the current maps) is shown at the left. It has 1000 randomly placed `deflections' 
or unreal responses (blue dots), and the primary beam is aimed at a radio population with a 
power-law source count of $N=KS^{-\gamma}$, 
where  $\gamma=1.5$. We supplied 10000 sources above 1~Jy in the total field of 204.8$^2$~arcsec$^2$. 
We set the detection limit at the centre as 1.0~Jy; the beam adopted
is Gaussian with FWHM = 60~arcsec. Selecting randomly from the source count gives the set of red dots,
falling off rapidly away from the centre as equation~\ref{radial} indicates. The right diagram shows 
the resulting radial distribution in histogram form (blue; the histogram of random background
deflections alone is red). The radii were chosen so that equal areas are encompassed 
by all annuli, i.e. $r_j = \sqrt(C/\pi + r_{j-1}^2)$. The models shown (green curves), calculated from 
equation~\ref{radial} with the 1000 uniformaly-random deflections added, have $K$ running from 
0.18 (lower) to 0.28 (upper); there is a clear selection at $K=0.23 \pm 0.02$ from the 
minimum chi-squared test. The value of $K$ we anticipate is $10000/204.8^2 = 0.238$ sources per unit 
area at a flux density of 1~Jy.}

\label{sim}

\end{figure*}

Consider an integral count of power-law form (and over the short flux range sampled, this is a
perfectly adequate description), $N(>S) = K S^{-\gamma}$, and a round beam of Gaussian shape 
\begin{equation}
f(r) = {\rm exp}[-4 {\rm ln}2r^2/\Delta^2], 
\end{equation}
where $\Delta$ is the FWHM of this Gaussian (i.e. the variance for the Gaussian is
$\sigma^2 = \Delta^2/8{\rm ln}2$). If we can detect a source of $S_0$~Jy at the centre of the beam, then 
the source density here is $N(r=0) = K S_0^{-\gamma}$; but if it is away from the centre, the source 
density must fall off as 
\begin{equation}
N(r)=K (S_0 / {\rm exp}[-4{\rm ln}2 r^2/\Delta^2])^{-\gamma}
\label{radial}
\end{equation}
If $\gamma = 1$ then the source detection level is $\propto$~(1/beam factor); but if $\gamma$ exceeds 
$1$, taking say the Euclidean value of $1.5$, then the falloff in source density is more rapid 
than the beam shape implies. Fig.~\ref{sim} shows how this can be used to estimate $K$. Note that 
{\em no values of flux density need to be measured} (except to set the survey limits). The result 
depends on a radial counting of source numbers alone. This is an illustrative approach; we now describe
a greatly improved likelihood technique.

%Suppose we try this test imply by adding all the fields together, ie just looking at the distribution
%of the 54225 'detections'. Fig.~\ref{fld_add} shows the result, and the best fit model. We have no idea
%here what the units of $K$ are; but it shows that a definite upper limit can be obtained. We already have
%excellent priors from \citet{wald07} to contrain the values of $K$ and this demonstrates that if we do
%the minimization in detail - a 1D approach using the \citet{mar83} technique, and summing the results for
%field by field - then we can get an estimate of $K$ with known units and applying the priors, 
%get definition of a region in the $N-S$ diagram at values of $S$ far below previous estimates.

%\begin{figure}
%\vspace{5.4cm}

%\special{psfile=real_fld_1_1.ps hoffset=0 voffset=180 vscale=31 hscale=31
%angle=-90}

%\caption{The distribution of all 54225 `deflections' found in the 1627 survey fields above $3\sigma$. Note 
%that the value of $\sigma$ varies between snapshots from 0.8 to 3~mJy, and hence there is no single survey
%limit and no way of estimating the area so that $K$ can be quoted in units of say sr$^{-1}$. The best fit 
%value of this non-normalized $k$ is $0.0024 + 0.0010 - 0.0024$, i.e. an upper bound on the source density.}

%\label{fld_add}

%\end{figure}

To set this up as a Maximum Likelihood problem, suppose first that there are
no fake sources, no noise-simulated source-like deflections, and the 
distribution of flux densities is given by an integral source
count law of the form $N(>S) = K S^{-\gamma}$. We choose annuli, radial elements,
small enough that their occupancy is either no deflection+sources, or one at maximum.
The $\cal L$(ikelihood) function for the $i^{\rm th}$ object
is the probability of observing {\it one} object in its
$r_i$ element times the probability of observing {\it zero} objects in all other $r_j$ elements 
accessible to it. The Poisson model is the obvious one for the likelihood:
\begin{equation}f(x:\mu)=\frac{{\rm e}^{-\mu}\mu^x}{x!},\end{equation}
where $\mu$ is the expected number. If $x=1$, the function is $\mu {\rm e}^{-\mu}$ 
and if $x=0$ it is ${\rm e}^{-\mu}$.

The expected number $\mu$ as a function of radial distance $r$ igiven by
\begin{equation}
\mu=\lambda(r)2 \pi r dr,\ \ \mbox{with}\ \
\lambda=KS^{-\gamma}dS, 
\end{equation}
\noindent and
\begin{equation} S=f(r),\ \mbox{where}\ f(r)=(S_0 / exp[-4{\rm ln}2r^2/\Delta^2])^{-\gamma}.
\end{equation}

\noindent To avoid arithmetic, we divide the area radially into one central circle and $N-1$ annuli
about this, each of area $2 \pi r dr = a_i$, and we design all the 
radius elements and the central circle to have the same area $a_i=a$ (see caption to Fig.~\ref{sim}). 
We get $\mu(r)=K f(r) a$, so that for a total of 
$n$ sources ($N>n$), 
\begin{equation}
{\cal L} = \prod_{i}^{N} \lambda(r_i) a
\,{\rm e}^{-\lambda(r_i)a}\,\prod_{j \neq i}^{N} {\rm e}^{-\lambda(r_j) a}\!,
\end{equation}

\noindent where $i$ denotes the element of the $r$ plane in
which sources are present and $j$ denotes all others. From this, if
$s=-2\,\ln \cal{L}$, then 
\begin{equation}
s = -2 \sum_{i=1}^n\,\ln f(r)_i - 2N \mbox{ln} (Ka) + 2Ka\sum_i^N f(r)_i .
\end{equation}

As a check at this point, if we set
the derivative of $s$ with respect to $K$ to zero, we get a
maximum-likelihood estimate for $K$:
\begin{equation}
    K_0 = \frac{n}{a\sum_i^N f(r)} = \frac{n}{a\sum_i^N [S_0/exp(-4 {\rm ln} 2 r^2/\Delta^2]^{-\gamma}}.
\end{equation}

\noindent Moreovoer if $\Delta=\infty$, i.e. an infinitely broad beam, and we set $Na = A$, the total area surveyed,
this reduces to $N = K_0 \, S_0^{-\gamma} \,A$,
i.e. the number of sources observed is the source-count per unit area times the area surveyed.

We now add a random uniform background of fake `sources', at a density of $C$ sources per unit
area. (Note that there is no assumption of an intensity distribution for these; this likelihood 
technique depends
solely on surface density.) We get the  value of $\lambda=KS^{-\gamma}dS + C$, 
so that $\mu = a(KS^{-\gamma}dS + C)$. With the same analysis as above, we find
\begin{eqnarray}
s & = & -2\sum_i^n {\rm ln}(Kf(r)_i + C) - 2(n{\rm ln}(a) +aCN) \nonumber\\
&&+ 2Ka\sum_i^Nf(r)_i.
\end{eqnarray}
There is now no simple differentiation to find a best estimate for $K_0$. Of course, setting $C = 0$
recovers the results of equations 7 and 8.

\begin{figure}
\vspace{6.cm}

\includegraphics{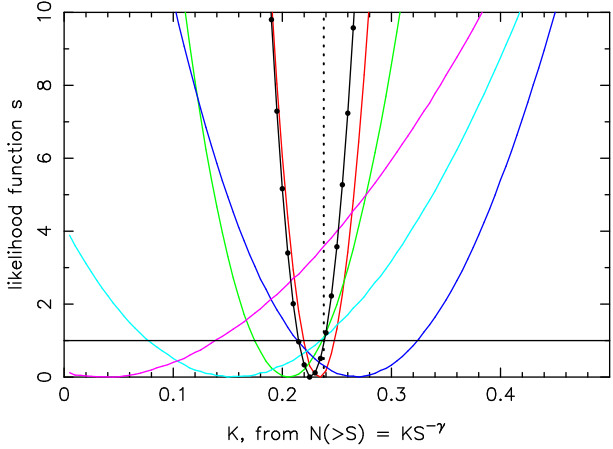}

\caption{This simulation adopts the original simulation parameters, but now selects 5 fields 
with the same initial source count. The random uniform-background samples added to all five fields 
have the same surface density, but the five have different cutoff-sensitivity
($S_0$) values of 1.0, 2.0, 3.0, 4.0 and 5.0 Jy.  Randomly-selecting from
the original source count via the 60-arcsec beam results in samples of 595, 
212, 111, 80 and 66 sources respectively, the number decreasing as the survey sensitivity
is decreased. The results of the likelihood calculation are shown by the curves of 
likelihood function $s$, red for 1.0 Jy, green for 2.0, blue for 3.0, turquoise for 4.0 and 
magenta for 5.0. The function broadens rapidly as we lose sources with reduced 
sensitivity -- see text. The black curve is the sum; the black line at $\Delta s = 1.0$ 
indicates the approximate $\pm1\sigma$ value so that the $\pm 1.0 \sigma$ range of K for
each of the 5 realizations (and for the summed black curve) is given by the range of $K$ between
the intersections with this line. The derived value of $K$ from the summed (black) curve is 
is $0.23$ cf the true value of 0.238.}

\label{svsk}

\end{figure}

The only point we now need to address is how to do this for 1627 fields combined, the problem
being that each field has a different value of $S_0$, the detection limit, because each
field has a different value of rms noise. The path towards this is straightforward, as 
shown in Fig.~\ref{svsk}. This simulation models the real-life situation by taking five 
simulations each using the same source count and beam parameters, and choosing five 
different values for detection levels 
$S_0$, at 1.0, 2.0, 3.0, 4.0 and 5.0 Jy. Of course higher values of $S_0$ yield far fewer
detections of sources. With the same number of randoms in each field, the likelihood 
function thus broadens rapidly with increasing $S_0$. Likelihoods are multiplicative, so 
that only addition (and normalization) of the likelihood functions computed for all fields
is required. The anticipated value of $K$ is retrieved as shown.

%Table~\ref{tab_svsk} shows the estimates s of $K$ obtained, and the $\pm 1\sigma$ (68 percent)
%range for each $K$ which is expected to encompass the true value.

\begin{figure}
\vspace{5.5cm}

\includegraphics{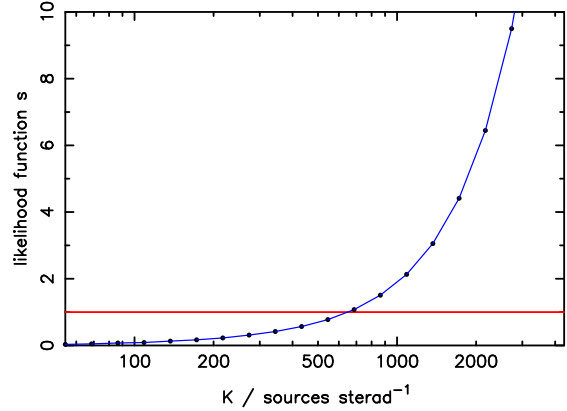}

\caption{The run of likelihood function $s$ vs $K$, the source-count normalization, 
summed for the 74 batches of 22 fields each at the points shown. There is no minimum; the imprint 
of the beam in the data is too weak to be detected, so that the likelihood function provides only an upper
limit to the source-count normalization $K$. The intersection of horizontal line at $s=1$ 
indicates the value of $K~(=652.5$) corresponding to a 1$\sigma$ (68 per cent) upper limit.}

\label{likfn}

\end{figure}

To perform the experiment on the results from the survey, we again reduced the total number of sources 
by limiting the total list of $>3.0\sigma$ detections to those which could be point-like. We adopted
the slightly more stringent limits of $0.6<{\rm SF}<1.4$, yielding a total of 26090 `sources'. Plotting
these in a histogram analogous to Figure~\ref{sim} showed no indication of an increased
surface density at the smaller radii, a clear indication that the likelihood calculation will yield only
an upper limit to $K$. Because individual fields have very small numbers of objects (mean = 26090/1627 
or $\sim 16$) on which to perform the likelihood calculations, we ordered the fields according to their 
rms values and added them together in 74 batches of 22. We then used the mean value of the rms=$\sigma$ 
for each batch of 22 to calculate a single value of $S_0 = 3.5\sigma$ for each group of 22 fields. 
We adopted a 60 arcsec beam and a value of $\gamma = 1.15$ from the estimate of Waldram et al. (2007). We 
calculated the likelihood function for each of the 74 batches for a range of values of $K$, and summed 
the results. The final curve of $s$ vs $K$ is shown in Figure~\ref{likfn}.

\begin{figure}
\vspace{11.8cm}

\includegraphics{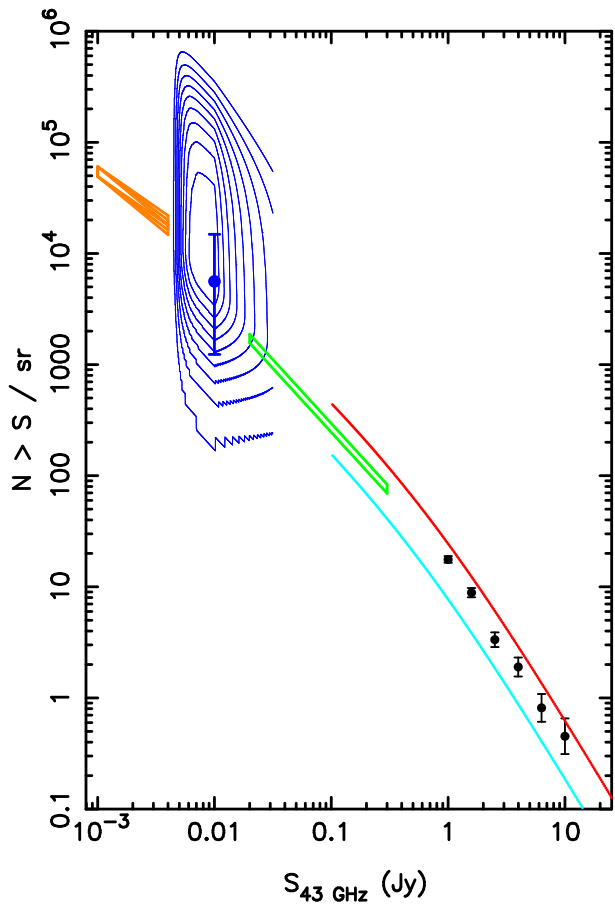}

\caption{An integral source count at 43 GHz. The blue contours outline a region of 
probability derived from the VLA 43~GHz survey as described in the text; they are at 
$p=0.1, 0.2 \ldots 0.9$. The single blue point at 10~mJy is from the one certain source 
detected in the survey. The black points at the high flux densities are from the WMAP 
all-sky survey (Wright et al. 2009). The green region describes the prediction at 43~GHz 
by Waldram et al. (2007) based on spectra and the 9C survey at 15 GHz. The red curve is the 
parametric representation of the AT20G counts at 20~GHz, and the turquoise curve shows 
the parametric form of the count estimated at 95~GHz (Sadler et al. 2008). The orange region  
describes the 31-GHz count estimate by Mason et al. (2009) from GBT measurements at this 
frequency of a sample of sources from the NVSS 1.4-GHz catalogue. No scaling has been 
applied to these or the 20-GHz and 95-GHz results from Sadler et al. (2008). The only direct 
measurements at 43~GHz are from the present set (blue contours and point) and the WMAP 
all-sky survey (black points).} 

\label{sc43}

\end{figure}

When converted to probabilities, this curve allows the construction of an integral source count 
probability region at 43 GHz (Figure~\ref{sc43}) in the following way. Because the likelihood
function curve (Fig.~\ref{likfn}) has no minimum, i.e. the imprint of the beam function is
not seen against the background of (almost completely) spurious `sources', the likelihood
calculation only provides an upper limit to $K$. Translating the likelhood function into
probabilities yields the upper-bound probability contours in Fig.~\ref{sc43}. As for the lower-limit
probability contours, we know that the 
survey found a minimum of one source (16382+414). Knowing the total area covered  
(Figure~\ref{area_mJy}), from Poisson statistics we can 
compute a lower-limit set of probability contours for surface density. The range over which 
our estimate of the source count pertains can be set at the high flux-density end with recourse to
the source count estimate itself, taking 100 per cent as the probability of finding one
source at the flux density of 16382+414 and adopting the (integral) slope value of -1.15. 
At the low-intensity end, we formed the distribution of values of 
$\sigma$ for all 1627 fields and assumed a detection threshold of $3.5\sigma$ to calculate a 
histogram of detection-level probability. 

Also shown in Fig.~\ref{sc43} is the 
43-GHz count estimate of Waldram et al. (2007), the counts from the WMAP 43-GHz source list 
(Wright et al. 2009), 
and the parametric forms of counts at 20 GHz and 95 GHz, as presented by Sadler et al. (2008). We also 
include the results from Mason et al. (2009) at 31 GHz, an integral source count estimate in the 
range 1 to 4~mJy of $N(>S) = (16.7 \pm 1.7) (S/1\,{\rm{mJy}})^{-0.80 \pm 0.07}$ deg$^{-2}$. It
must be emphasized that of the data in this diagram, only our 43-GHz point and contours plus the
WMAP source count represent direct observations at 43~GHz. The other direct observational result
in the diagram is the representation of the 20-GHz souce count from the AT20G survey. All other
data are inferred or projected from 31~GHz, 43~GHz or 95~GHz. There is no scaling of any result in the diagram.

\section{Discussion and Conclusions}
\label{disc}
1. This 43-GHz survey (VLA D-array) reaches approximately 7 mJy over an area of 0.5 \sdeg, with a 
primary beam of 1.0~arcmin FWHM and a circular synthesized beam of 2.3~arcsec FWHM. Done in snapshot mode
to follow the `wide-shallow' dictum, it found many apparent sources, most of which on inspection proved 
to be at the junction of response lines. Most can be eliminated on this basis, and on the basis that they 
are extended to the 2.3 arcsec synthesized beam. The radial distribution likewise makes it evident that 
the overwhelming majority of the deflections are not real sources.

2. However the survey does detect real sources, the flat-spectrum QSO 16382+414 for certain, and some
additional not-very-probable candidates (Table~\ref{coinc_all}) for which further observations are 
required. One of these is coincident in position with previously-catalogued sources at 1.4~GHz; there is no
candidate identification above the limits of the SDSS. Of the 6 other `sources' in the table, all are
coincident with blue stellar objects close to the region occupied by $z \sim 2.5$ FSRQ. These putative
6 sources all must have extreme inverted spectra because stacking suggests that the average flux density
at 1.4~GHz must be below 0.1 mJy.

3. We have developed a likelihood method to estimate source counts in a synthesis survey which has 
thrown up a combination of real and random sources. The method examines the radial distribution in 
each synthesized field, and because the real sources should decrease with radial distance from the 
field centre in a predictable 
manner, in principle both the source-count normalization and the source count power-law index can be 
determined. With this process, neither detailed cataloguing of the sources nor {\it a priori} 
decisions about which deflections are real are necessary. We have applied this method to the 
present data, but the dominance of random deflections (by a ratio of perhaps 10000:1) is such 
that only an upper limit to the surface density of real sources can be derived. No information 
on the slope of the count is available from the present data; we have adopted a slope estimate 
for the appropriate flux-density range from Waldram et al. (2007), although the resulting upper-limit 
contours are relatively insensitive to this. 

4. The present results are consistent with extrapolating the count estimate of Waldram et al. (2007), and 
with the 31-GHz count estimate of Mason et al. (2009), but suggestive of a somewhat higher 
source surface density than found in either investigation. Such a surface density enhancement might
account for the excess signal at $\ell > 2000$ found by CBI. We note that the De Zotti et al. (2005) 
count models, calculated at high frequencies from detailed synthesis of populations from lower-frequency 
surveys, lie significantly above the count estimates of Waldram et al. (2007).

5. In terms of individual sources, the 43-GHz survey yield is small and provides no {\it direct} 
evidence that any new types of inverted-spectrum sources appear at mJy levels. The lower bound of our 
error box in Fig.~\ref{sc43} is strongly dependent on the certain detection of a
single source; if any of our candidate sources of Table~\ref{coinc_all} should prove to be real, then
a count enhanced above the Waldram et al. (2007) estimate becomes probable. Any such sources will have
extremely inverted spectra, as we and Mason et al. (2009) have noted. 

6. The observations indicate that there are considerable challenges involved in a snapshot synthesis 
survey at a frequency as high as 43 GHz.

\vspace{0.5cm}

\noindent {\bf Acknowledgements} We thank Chris Willott for helpful discussions about field choice. 
We are very grateful to Barry Clark for his efforts in scheduling the VLA to accommodate a late 
improvement in our observing proposal. Jim Condon proferred useful advice.
Eric Greisen gave much help and advice with running AIPS and checking
systems remotely. Melanie Gendre helped with running some of the AIPS procedures. Ron Ekers 
offered many constructive comments on an early draft. Chris Blake gave useful advice on SDSS data. 
We thank the referee for helpful comments.
JVW would particularly like to thank Katherine Blundell for help with computational issues and 
and with running AIPS over the past several years. JVW acknowledges support during the course 
of this work through Canada NSERC Discovery Grants.

The National Radio Astronomy Observatory is a facility of the National Science Foundation operated 
under cooperative agreement by Associated Universities, Inc.

This research has made use of data obtained from the SuperCOSMOS Sky Surveys
(SSS), prepared and hosted by the Wide Field Astronomy Unit, Institute for
Astronomy, University of Edinburgh, which is funded by the UK Science and
Technology Facilities Council. The SSS Web Site is {\it www-wfau.roe.ac.uk/sss/}.

The research also made use of data from the SDSS survey. Funding for the SDSS and SDSS-II has been 
provided by the Alfred P. Sloan Foundation, the 
Participating Institutions, the National Science Foundation, the U.S. Department of Energy, the 
National Aeronautics and Space Administration, the Japanese Monbukagakusho, the Max Planck 
Society, and the Higher Education Funding Council for England. 
The SDSS Web Site is {\it www.sdss.org/}.

\appendix

\section{Observing strategy}

Given an integral source-count of the form $N = K S^{-\gamma}$, integration time
$t$ per pointing, and a telescope settle time $t_s$, there is an optimum observing
strategy to detect the maximum number of sources.

It is easy to show that in the radio regime the {\it maximum number of pointings} is 
optimal; wide and shallow easily beats deep and narrow. In a given time T we can 
make $T/(t+t_s)$ pointings. Consider the case for long integrations in which $t_s$ is negligible.
Then the number of pointings is T/t, and if the equivalent beam area is $A_o$ then
the area covered is $A=A_oT/t$. The flux density limit becomes $S=S_{o} /
\sqrt{t}$, where $S_o$ is the flux density reached in unit time. The total number
of sources found is $N = AK S^{-\gamma} = T/t (S_{o} /\sqrt{t})^{-\gamma}$, or $N
\propto T t^{(\gamma/2 - 1)}$. Thus until $\gamma > 2$, until the source count is
{\it steeper} than that even at highest radio flux densities, the number of sources is
maximized with minimum integration times and maximum number of pointings. In fact
at the flux densities in question $\gamma \sim 1$.

Given then that short exposures are required, what is optimum exposure time, in
that $t_s$ must now come into play? There must be an optimum as if we let $t
\rightarrow 0$, $t_s$ in itself imposes a maximum number of pointings for {\it
zero} integration time. So let $t = xt_s$. The limiting flux density reached is $S
= S_o/\sqrt{t}$; the area covered in given observing time $T$ (in which $T/t(1+x)$
pointings are possible) is $A_o T / t_s(1+x)$. The total number of sources
detected will be

\begin{equation}
N(x) = \frac{TAKS_o^{-\gamma}}{t_s^{1.5\gamma}} \frac {x^{\gamma/2}}{1 + x}
\end{equation}
\noindent maximized at
\begin{equation}
x_{\rm max} = \frac{\gamma}{2 - \gamma}
\end{equation}

\begin{figure}
\vspace{5cm} \includegraphics{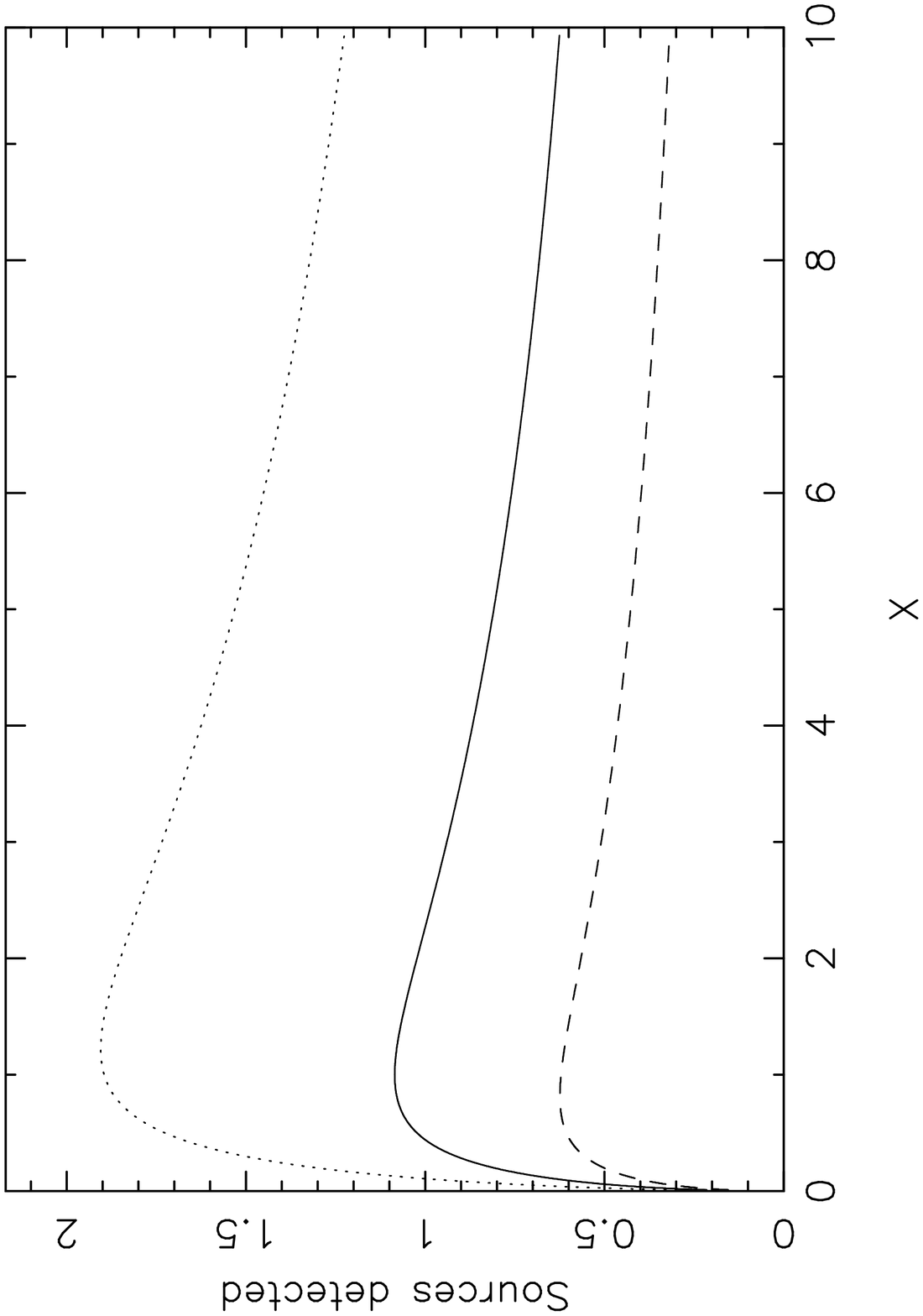}

\caption{The function of equation~1; the source yield as a function of the ratio
$x = t/t_s$ where $t$ is integration time per snapshot and $t_s$ is the telescope
settle time. For an integral source count of the form $N=KS^{-\gamma}$, the dashed
curve represents $\gamma = 0.9$, the solid curve $\gamma = 1.0$ and the dotted
curve $\gamma = 1.1$.}

\label{fig2_1}
\end{figure}

\noindent For $\gamma = 1$ the result is particularly simple : $x_{\rm max} = 1$,
integration time and settle time should be equal. Fig \ref{fig2_1} shows the
function of equation~1. At $x \ge x_{\rm max}$ the function falls very slowly
indeed. This insensitivity to $x$ suggests that a further factor could be
considered - how much data do we not need? For $x = 3$ we achieve only half the
pointings as for $x=1$, and yet the loss of sources is just 15~per~cent. The NVSS
used a 30s cycle; 23s integration with 7s settle, $x = 3.3$, representing a good
compromise.

The final factor to consider is whether to overlap the beams to make the survey
complete to a limiting flux density or to have them entirely independent in a
coarse grid. The argument of the opening paragraph pertains - the answer is to
use completely independent beams to yield maximum area at minimum sensitivity. The
issue can be quantified, quoting the analysis communicated to us by Jim Condon.
The NVSS spacing is overlapped just enough to provide nearly uniform sensitivity
to ensure {\it completeness} down to a fixed limit (e.g., 2.5 mJy/beam) over areas
$>>$ beam solid angle.  To maximize the number of detections in a {\it representative} 
sample without achieving completeness, the optimum spacing is
anything large enough that successive snapshots have no overlap (i.e. $> 2$ FWHM).
Then, instead of a survey with uniform sensitivity, the yield is a survey covering
a small area with high sensitivity and larger areas with lower sensitivity.  The
NVSS spacing gives the same sensitivity as the on-axis sensitivity of a single
snapshot over an effective area of 1/2 the beam solid angle, or $\pi \theta^2 / (8
{\rm ln} 2)$ for a Gaussian beam with FWHM $\theta$ (Condon et al. 1998).  A single isolated
snapshot has sensitivity directly proportional to the primary power pattern.
Since the source counts are roughly power-law with a differential slope of $2.0$
(integral slope of $1.0$), the number of detectable sources in the isolated
snapshot will be proportional to the whole beam solid angle $= \pi \theta^2 / (4
{\rm ln} 2)$.  Thus the yield per unit time with non-overlapping snapshots is 
{\it twice} that with snapshots overlapped as for the NVSS gridding.

Accordingly we adopted the NVSS 30s cycle, 7s settle and 23s integration, and
coarse gridding. (A shorter settle time for the VLA would have been an advantage to us.)

%\bibliographystyle{aa}
%\bibliographystyle{apj}
%%\bibliographystyle{mn2e}

%\begin{thebibliography}{99}

%\bibliography{qband_bib}
%%\bibliography{qband_bib_2}

%\begin{thebibliography}{}
%\end{thebibliography}

%\begin{thebibliography}{}
%\end{thebibliography}

\end{document}